\documentclass[aps,groupedaddress,showpacs]{revtex4}
\usepackage{graphicx}
\begin{document}

\title{Nucleation of superconductivity in Al mesoscopic disk with magnetic dot}
\author{ D. S. Golubovi\'{c}, W. V. Pogosov, M. Morelle, and V. V. Moshchalkov}
\affiliation{Laboratorium voor Vaste-Stoffysica en Magnetisme,
Katholieke Universiteit Leuven, Celestijnenlaan 200 D, B - 3001
Leuven , Belgi$\ddot{e}$}

\begin{abstract}
We have studied the nucleation of superconductivity in a
mesoscopic Al disk with a Co/Pd magnetic dot placed on the top by
measuring the normal/superconducting phase boundary $T_{c}(B)$.
The measurements have revealed a pronounced asymmetry in the phase
boundary with respect to the direction of the applied magnetic
field, indicating an enhancement of the critical field when an
applied magnetic field is oriented parallel to the magnetization
of the magnetic dot. The theoretical $T_{c}(B)$ curve is in a good
agreement with the experimental data.
\end{abstract}

 \maketitle

Superconductivity at the sub-micrometer scale has been extensively
studied over the past decade (e. g. reference \cite{victor} and
references therein). Recently, hybrid mesoscopic
ferromagnetic/superconducting systems have attracted a
considerable attention, since it is believed that the interaction
between ferromagnetism and superconductivity at the mesoscopic
scale may lead to a number of new physical effects \cite{misko,
miskol,mjvb,martin,erdin}. In addition, these structures are
considered a prominent candidate for technological applications as
they offer a unique possibility of tuning the field range in which
superconductivity nucleates.

A mesoscopic disk in a nonuniform magnetic field, resembling a
magnetic dot with the perpendicular magnetization or a current
loop, has recently been studied theoretically by using the
Ginzburg-Landau equations \cite{misko}. We report results on an
individual hybrid mesoscopic ferromagnetic/superconducting
structure, consisting of an Al disk with a Co/Pd magnetic dot on
the top. The magnetic dot has a perpendicular magnetization. %The
%superconducting phase boundary $T_{c}(B)$ of the hybrid structure
%has been determined by four-point transport measurements in a
%cryogenic setup applying magnetic fields perpendicularly to the
%sample surface.
The measurements have revealed a pronounced asymmetry in the
superconducting $T_{c}(B)$ phase boundary with respect to the
direction of a perpendicular applied magnetic field.

The sample was prepared on a SiO$_{2}$ substrate, by electron beam
lithography on a PMMA950K and co-polymer electron beam resists in
three steps. In each step, a desired structure was patterned, the
material(materials) evaporated and the final structure obtained
with a  lift-off in warm acetone. The thickness of the Al disk is
$450\,$\AA $\,$, whereas the magnetic dot consists of a
$25\,$\AA$\,$ Pd buffer layer and ten Co/Pd bilayers with the
thicknesses $4\,$\AA$\,$ and $10\,$\AA, respectively. The magnetic
dot is separated from the disk by a layer of Al oxide, which forms
nearly instantly on the surface when the structure is exposed to
the air.
\begin{figure}[htb]
\centering
\includegraphics*[width=8.5cm]{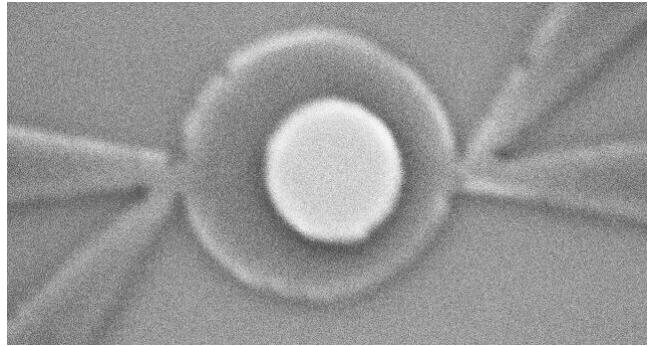}
\caption{A scanning electron micrograph of the structure.
\label{fig1}}
\end{figure}

Fig. \ref{fig1} shows a scanning electron micrograph of the
structure. A bright area in the image is the magnetic dot. The
radii of the disk and magnetic dot are $1\,{\rm \mu m}$ and
$0.5\,{\rm \mu m}$, respectively. The disk has wedge-shaped
contacts with the opening angle of $15^{\circ}$ since these have
proved to be the optimum for transport measurements of mesoscopic
superconducting structures \cite{proka}. The magnetic dot is
displaced, approximately by $130\,{\rm nm}$, from the center of
the disk, despite a painstaking alignment procedure. We believe
that, in addition to the limitations of the electron beam writer
at our disposal, a minor displacement of the  alignment markers
may have contributed to the shift of the magnetic dot.

\begin{figure}[htb!]
\centering
\includegraphics*[width=8.5cm]{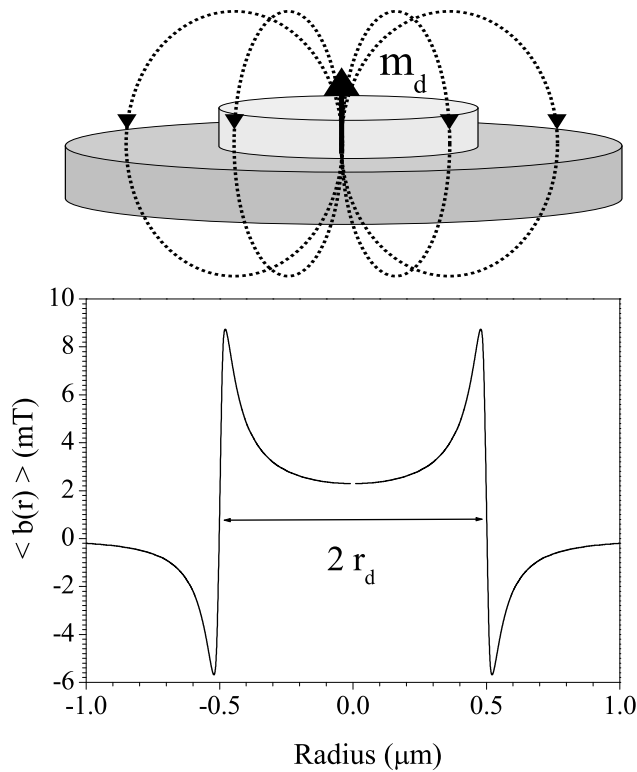}
\caption{The calculated stray field of Co/Pd magnetic dot with the
radius of $r_{d}=500\,$nm and height $h_{d}=16.5\,$nm, averaged
over the thickness of the Al disk.\label{mag}}
\end{figure}

The chosen composition of Co and Pd provides perpendicular
magnetization of the magnetic dot, as confirmed by the
magneto-optical Kerr measurements of the co-evaporated reference
film at room temperature, which revealed a full remanence and the
coercive field of $350\,{\rm mT}$. For a thorough analysis of the
properties of patterned Co/Pd structures we refer to
\cite{martin}. Fig. \ref{mag} shows the calculated stray field of
the magnetic dot, averaged over the thickness of the
superconducting disk $<b(r)>$ for the radius and height of the dot
$r_{d}=500\,{\rm nm}$ and $h_{d}=16.5\,{\rm nm}$, respectively.
The brighter area in the schematic is the magnetic dot, darker
area is the superconducting disk whereas the arrows indicate the
field distribution.

Prior to the measurements, the magnetic dot was magnetized
perpendicularly in a magnetic field of $450\,{\rm mT}$.
Superconducting phase boundary $T_{c}(B)$ was obtained from
four-point transport measurements, using an ac current with the
rms of $100\,{\rm nA}$ and frequency $27.7\,{\rm Hz}$, in a
cryogenic setup with the temperature stability of $0.2\,{\rm mK}$.
The phase boundary was measured resistively by sweeping magnetic
field at a very slow rate at a constant temperature and making use
of a lock-in amplifier in order to improve the signal-to-noise
ratio.

\begin{figure}[htb]
\centering
\includegraphics*[width=8.5cm]{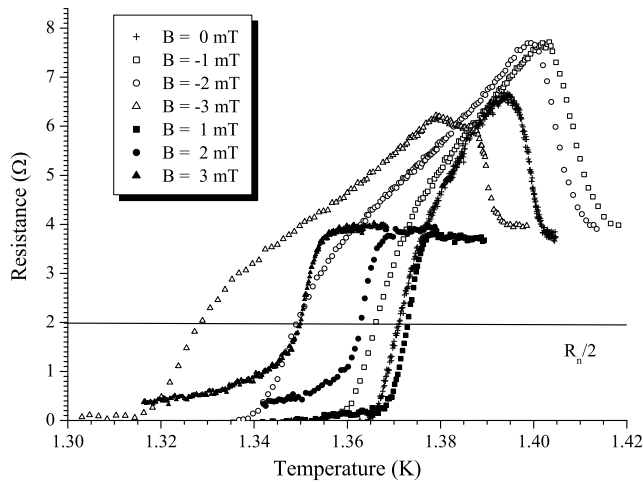}
\caption{Resistive transitions $R(T)$ for zero and $\pm 1\,$mT,
$\pm 2\,$mT and $\pm 3\,$mT applied fields. The curve with crosses
is the resistive transition in zero field, filled curves are the
resistive transitions when an applied field and the magnetization
of the magnetic dot are parallel (positive fields), whereas the
open curves indicate resistive transition for antiparallel
orientation (negative fields).\label{fig2}}
\end{figure}

Fig. \ref{fig2} presents the resistance of the structure versus
temperature measured in a constant external field. The curve with
crosses is the resistive transition in zero applied field, filled
curves correspond to the case when the magnetization of the dot
and external fields are parallel whereas open curves give the
resistive transition for the antiparallel orientation. The values
of the applied fields were $\pm \, 1\,{\rm mT}$,$\pm \, 2\,{\rm
mT}$ and $\pm \, 3\,{\rm mT}$, respectively. Plus signs indicate
parallel and minus signs antiparallel orientation. This convention
will be used throughout the paper.

The horizontal line indicates the resistive criterion $R_{n}/2$ (
$R_{n}=4\,{\rm \Omega} $ is the resistance in the normal state )
that was used for the determination of the phase boundary. There
is a clear asymmetry between the resistive transitions for the
same value but different orientation of an applied magnetic field
with higher critical temperatures for positive applied fields.
Moreover, when the applied magnetic field is $1\,{\rm mT}$, the
critical temperature is slightly higher than the critical
temperature in zero field, which is a clear evidence that for the
particular parameters of the disk and magnetic dot, this external
magnetic field provides an enhancement of superconductivity in the
disk. The enhancement demonstrates that
ferromagnetic/superconducting mesoscopic structures make it
possible to tune the field range in which superconductivity
nucleates.

Since the magnetic dot generates a finite flux through the disk in
zero field, its critical temperature is effectively decreased when
compared to the mesoscopic contacts and superconductivity
nucleates nonuniformly. As a result, the resistive transition in
zero applied field is broad and has a temperature dependent slope.
When increasing temperature first the disk becomes normal, and the
slope of the transition, now mainly determined by the mesoscopic
contacts, changes. When a uniform positive magnetic field is
applied the effective magnetic induction in the disk locally
increases in the region below the dot, whereas in the rest of the
disk decreases. For a negative uniform field the situation is just
the opposite. Given the intensity of the stray and applied fields,
it is clear that the negative applied fields further suppress
superconductivity and decrease the critical temperature of the
disk thus causing broader transitions. Accordingly, the slope of
the transitions changes at a lower temperature as the negative
applied field increases. The transitions for the positive applied
fields are steep, and more importantly, steeper then the
transition in zero field, due to the reduction of the local
magnetic induction in the disk around the magnetic dot and,
consequently, an increase in its critical temperature. The tails
in the transitions are a result of the flux flow.

The transitions for zero and the negative applied fields exhibit a
considerable overshoot in the resistance with respect to the
resistance in the normal state, nearly $100\,\%$ for applied
fields of $-1\,{\rm mT}$ and $-2\,{\rm mT}$. This effect comes
about due to the charge imbalance effects at a
superconductor/normal metal junction, %In mesoscopic
%superconducting structures the charge imbalance effects are
caused by a geometrically imposed difference in the critical
temperatures of the structure itself and the contacts, or a local
suppression of superconductivity in the narrowest part of the
contacts by a transport current. For the details we refer to
\cite{proka}. When there is a considerable difference in the
critical temperature of the disk and mesoscopic contacts, that is,
for zero and negative applied fields, overshoots appear in the
resistive transition.

The order parameter is strongly suppressed at the
superconductor-normal metal boundary due to the proximity effect
\cite{de Gennes}. In spite of the oxide layer on the disk, Co/Pd
dot suppresses superconductivity inside the disk below the
magnetic dot. Therefore, the disk with the magnetic dot may be
approximated by a superconducting loop of a finite width. Note
that we have also studied experimentally the nucleation of
superconductivity in an Al disk with the radius of $800$ nm and
the same magnetic dot, but were unable to drive the system to the
superconducting state. We believe that this is a result of the
suppression of the Cooper pair density by the contact with the
metal. We have used a simple model assuming that the order
parameter is constant within the effective loop, and that the
inner radius of the loop equals the radius of the dot. %The value
%of the order parameter can be found by the minimization of the
%Ginzburg-Landau energy.
In the vicinity of the phase boundary, the local field in the loop
is equal to the sum of an applied field and the stray field,
whereas the modulus of the order parameter is cylindrically
symmetric, with the winding number (vorticity) $L=0$ representing
the Meissner state, $L=1$ single-vortex state and $L>1$
giant-vortex state \cite{victor}.

\begin{figure}
\centering
\includegraphics[width=8.5cm]{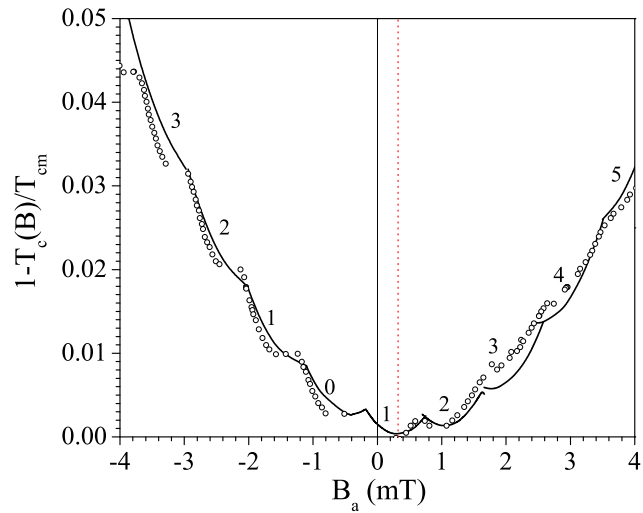}
\caption{Superconducting phase boundary $1-T_{c}(B)/T_{cm}$ versus
the applied field $B_{a}$. The open symbols show the experimental
phase transition, whereas the solid line presents the theoretical
results. \label{phase}}
\end{figure}

Fig. 4 shows the superconducting phase boundary, displayed as the
normalized critical temperature $1-T_{c}(B)/T_{cm}$ versus the
applied field. $T_{cm}=1.376\,{\rm K}$ is the maximum critical
temperature. Open symbols and solid lines correspond to the
experimental and theoretical results, respectively. The
theoretical data were obtained for the coherence length
$\xi(T=0)=76\,$nm and the outer loop radius which is $25\%$ bigger
than the real radius of the disk. This discrepancy can be
accounted for by the influence of the mesoscopic contacts, as well
as by the displacement of the magnetic dot from the center of the
disk. Even though a simplistic theoretical model has been used,
the agreement between the experimental data and theoretical curve
is good. The quasi-periodicity of the phase boundary is a result
of transitions from the normal state to the phases with different
vorticities $L$. Numbers in the Fig. \ref{phase} show the values
of $L$. Note that the magnetization of the dot is high enough to
create one vortex in the sample even in the absence of an external
field. The superconducting phase boundary strongly depends upon
the direction of an applied magnetic field, with higher critical
fields for the parallel orientation (positive applied fields) in
the whole range of investigation.

In conclusion, we have fabricated a mesoscopic superconducting
disk made up of Al with a Co/Pd magnetic dot on the top. The
superconducting properties of the system have been investigated by
measuring the superconducting/normal state phase boundary
$T_{c}(B)$. It has been demonstrated that the critical field is
higher when an applied magnetic field and the magnetization of the
magnetic dot are parallel. The experimental data are in a good
agreement with the theoretical $T_{c}(B)$ curve.

This work has been supported by the Belgian UIAP, the Flemish GOA
and FWO programmes, as well as by the ESF programme "VORTEX". W.
V. P. acknowledges the support by the Research Council of the K.U.
Leuven.


\begin{thebibliography}{10}
\bibitem{victor} V. V. Moshchalkov et al, {\em Handbook of
Nanostructured Materials and Nanotechnology 3}, ed. H. S. Nalwa,
(Academic Press, San Diego, 2000.)
\bibitem{misko} M. V. Milo\v{s}evi\'{c}, S. V. Yampolskii and F.
M. Peeters, Phys. Rev. B, {\bf 66}, 024515 (2002).
\bibitem{miskol}M. V. Milo\v{s}evi\'{c}, S. V. Yampolskii and F.
M. Peeters, Phys. Rev. B, {\bf 66}, 174519 (2002).
\bibitem{erdin} S. Erdin, I. F. Lyuksyutov, V L. Pokrovsky and V. M. Vinokur, Phys. Rev. Lett., {\bf 88}, 1 (2002).
\bibitem{mjvb} M. J. Van Bael, J. Bekaert, K. Temst, L. Van Look, V. V. Moshchalkov, Y.
Bruynseraede, G. D. Howells, A. N. Grigorenko, S. J. Bending and
G. H. Borghs, Phys. Rev. Lett., {\bf 86}, 1 (2001).
\bibitem{martin} M. Lange, M. J. Van Bael, Y. Bruynseraede and V. V. Moshchalkov,
Phys. Rev. Lett., {\bf 90}, 19 (2003).
\bibitem{proka}M. Morelle,Y. Bruynseraede and V. V.
Moshchalkov, Phys. Stat. Sol., {\bf 237}, 1 (2003).
\bibitem{vital} V. Bruyndoncx, L. Van Look, M. Verschure and V. V.
Moshchalkov, Phys. Rev. B, {\bf 60}, 14 (1999).
\bibitem{de Gennes} P. G. de Gennes, {\em Superconductivity of Metals and Alloys} (Benjamin, New York, 1966).


\end {thebibliography}
\end{document}